\documentclass[aps,floats,nofootinbib,amssymb,preprint,tighten,superscriptaddress]{revtex4}
\usepackage{epsf,epsfig}

\usepackage{graphicx}
\usepackage{color}

\newcommand{\be}{\begin{equation}}
\newcommand{\ee}{\end{equation}}
\newcommand{\bea}{\begin{eqnarray}}
\newcommand{\eea}{\end{eqnarray}}
\newcommand{\nn}{\nonumber}

\renewcommand{\check}{({\bf This statement needs to be checked!!})}

\newcommand{\half}{\frac{1}{2}}

\renewcommand{\slash}{\displaystyle{\not}}

\newcommand{\ba}{\begin{array}}
\newcommand{\ea}{\end{array}}
\newcommand{\bi}{\begin{itemize}}
\newcommand{\ei}{\end{itemize}}
\newcommand{\ben}{\begin{enumerate}}
\newcommand{\een}{\end{enumerate}}

\newcommand{\cs}{\mathbb S}

\preprint{
\hbox to \hsize{
\hfill$\vcenter{\hbox{\bf MADPH-08-1516}
                \hbox{\bf NUHEP-TH/08-06}
                \hbox{\bf ANL-HEP-PR-08-58}
		\hbox{\bf NPAC-08-21}
                \hbox{November 2008}}$}
}

\begin{document}
\title{\vspace*{.75in}
Complex Singlet Extension of the Standard Model}
\author{Vernon Barger}
\affiliation{Department of Physics, University of Wisconsin, Madison, WI 53706}
\author{Paul Langacker}
\affiliation{School of Natural Sciences, Institute for Advanced Study, Einstein Drive Princeton, NJ 08540}
\author{Mathew McCaskey}
\affiliation{Department of Physics, University of Wisconsin, Madison, WI 53706}
\author{Michael Ramsey-Musolf}
\affiliation{Department of Physics, University of Wisconsin, Madison, WI 53706}
\affiliation{Kellogg Radiation Laboratory, California Institute of Technology, Pasadena, CA 91125}
\author{Gabe Shaughnessy}
\affiliation{Department of Physics, University of Wisconsin, Madison, WI 53706}
\affiliation{Northwestern University, Department of Physics and 
Astronomy, 2145 Sheridan Road, Evanston, IL 60208 USA}
\affiliation{HEP Division, Argonne National Lab, Argonne IL 60439 USA}

\thispagestyle{empty}
\begin{abstract}
We analyze a simple extension of the Standard Model (SM) obtained by adding a complex singlet to the scalar sector (cxSM). We show that the cxSM can contain one or two viable cold dark matter candidates and analyze the conditions on the parameters of the scalar potential that yield the observed relic density. When the cxSM potential contains a global $U(1)$ symmetry that is both softly and spontaneously broken, it contains both a viable dark matter candidate and the ingredients necessary for a strong first order electroweak phase transition as needed for electroweak baryogenesis. We also study the implications of the model for discovery of a Higgs boson at the Large Hadron Collider. 
\end{abstract}
\maketitle

\newpage

\section{Introduction}

The Standard Model (SM) has been enormously successful in describing a plethora of electroweak and strong interaction phenomena, and many of its predictions, such as the existence of the top quark with a heavy mass as implied by electroweak precision data, have been confirmed experimentally. Nevertheless, the search for new physics beyond the SM has strong theoretical and experimental motivation. In this paper, we focus on the quest to explain the mechanism of electroweak symmetry breaking (EWSB) and the implications for two unsolved problems in cosmology: the nature of the non-baryonic cold dark matter (CDM) of the universe and the origin of the cosmic baryon asymmetry. The SM paradigm for EWSB, which relies on the Higgs mechanism with a single $SU(2)$ doublet, has yet to be confirmed, and the lower bound $M_H\geq 114.4$ GeV obtained at LEP II~\cite{Barate:2003sz} leads to some tension with the global set of electroweak precision observables (EWPOs) that favor a relatively light Higgs~\cite{Amsler:2008zz,LEPEWWG} with $M_H=84^{+32}_{-24}$ GeV~\cite{Renton:2008,Kile:2007ts}. From the cosmological standpoint, the identity of the CDM remains elusive, while the SM fails to provide the level of CP-violation or the strong first order electroweak phase transition (EWPT) that would be needed to explain the generation of baryon asymmetry during the EWSB era. 

Over the years, particle theorists have extensively studied a variety of specific scenarios for an extended SM -- such as the minimal supersymmetric standard model -- that address these questions.  It is possible, however, that the results of upcoming experiments at the CERN Large Hadron Collider (LHC) will not favor any of the conventional extended SM scenarios, leading one to consider new possibilities that will address the open problems at the cosmology-particle physics interface. In this paper, we consider a simple extension of the SM scalar sector that illustrates the necessary ingredients of such a theory. The simplest extension (xSM) entails the addition of a single, real singlet scalar to the SM scalar potential. The phenomenology of such a model has been analyzed in earlier work~\cite{McDonald:1993ex,Burgess:2000yq,O'Connell:2006wi,BahatTreidel:2006kx,Barger:2007im,He:2007tt,Davoudiasl:2004be}.  It has been shown that in the xSM the real scalar $S$ can either (a) provide a CDM candidate whose dynamics lead to the observed relic abundance, $\Omega_\mathrm{CDM}=0.1143\pm 0.0034$~\cite{Komatsu:2008hk}, or (b) lead to a strong first order EWPT as needed for electroweak baryogenesis, but not both simultaneously. Moreover, the latter possibility also allows for additional, light scalar contributions to the gauge boson propagators that alleviate the EWPO-direct search tension. In both cases, it is possible that the extended Higgs sector of the xSM could be identified at the LHC, and the discovery potential has been analyzed in detail in Ref.~\cite{Barger:2007im}. 

Here, we consider the next simplest extension of the SM scalar sector obtained with the addition of a complex scalar singlet field, $\cs$, to the SM Lagrangian (cxSM). We show that when the potential $V(H,\cs)$ has a global $U(1)$ symmetry that is both spontaneously and softly broken, it contains the ingredients needed to provide a viable CDM candidate, help generate a first order EWPT, and relieve the tension between the direct search bounds on $m_H$ and EWPO implications. We also analyze the conditions under which  cxSM dark matter yields the observed relic density and study the corresponding implications for the discovery of at least one cxSM scalar at the LHC.  In the absence of spontaneous symmetry breaking, the cxSM can give rise to a viable two-component dark matter scenario.  Either way, we show that a combined Higgs boson search that includes both traditional and ``invisible'' modes can enhance the LHC discovery potential for SM extensions with an augmented scalar sector.  Our analysis of the model is organized in the remainder of the paper as follows. In Section \ref{sec:vac} we discuss the potential and its vacuum structure, classifying the different possibilities for symmetry-breaking and summarizing the corresponding phenomenological implications. These possibilities are summarized in Table \ref{tab:cases}. Section \ref{sec:spectra} gives the spectra of physical scalar states for each of the scenarios in Table \ref{tab:cases}. In Section \ref{sec:const} we summarize the constraints on the model parameters implied by electroweak data, collider searches, and astrophysical considerations. Section \ref{sec:pheno} contains our analysis of the relic density and implications for Higgs discovery at the LHC. We summarize the main features of our study in Section \ref{sec:concl}.

\section{The cxSM and its Vacuum Structure}
\label{sec:vac}

The most general renormalizable scalar potential obtained by the addition of a complex scalar singlet to the SM Higgs sector is  given by
\bea
V(H,\cs) &=& \frac{m^2}{2}H^\dagger H+\frac{\lambda}{4}(H^\dagger H)^2+\left(\frac{|\delta_{1}|e^{i \phi_{\delta_1}}}{4} H^\dagger H \cs + c.c.\right)+\frac{\delta_2}{2} H^\dagger H |\cs|^2\nn\\
&+&\left(\frac{|\delta_{3}|e^{i \phi_{\delta_3}}}{4} H^\dagger H \cs^2+h.c\right)
+\left(|a_{1}|e^{i \phi_{a_{1}}}\cs+c.c.\right)+\left(\frac{|b_{1}|e^{i \phi_{b_1}}}{4}\cs^2+c.c.\right)\nn\\
&+&\frac{b_{2}}{2}|\cs|^2+\left(\frac{|c_1| e^{i \phi_{c_1}}}{6}\cs^3+c.c.\right)
+\left(\frac{|c_2| e^{i \phi_{c_2}}}{6}\cs|\cs|^2+c.c.\right)\nn\\
&+&\left(\frac{|d_1| e^{i \phi_{d_1}}}{8}\cs^4+c.c.\right)+\left(\frac{|d_3| e^{i \phi_{d_3}}}{8} \cs^{2}|\cs|^{2}+c.c.\right)+\frac{d_2}{4}|\cs|^4
\label{eq:pot}
\eea

where $H$ is the $SU(2)$ doublet field that acquires a vacuum expectation value (vev)

\be
\langle H\rangle = \left(
\begin{array}{c}
0 \\ v/\sqrt{2}
\end{array}\right) \ \ \ ,
\ee

and $m^{2}$ and $\lambda$ are the usual parameters of the SM Higgs potential.  The value of the SM vev we adopt is $v=246$ GeV.

The form of this potential is equivalent to one obtained by addition to the SM Higgs potential of two real scalar singlets, corresponding to the real and imaginary parts of $\cs$. For our purposes, however, it is convenient to work with the complex scalar. In addition, the essential features of the cxSM scenario can be realized after simplifying $V(H,\cs)$ through the imposition of two symmetries:

\begin{itemize}
\item[(a)] A discrete, $\cs\to -\cs$ (or $\mathbb Z_{2}$), symmetry may be imposed to eliminate all terms containing odd powers of the singlet field $\cs$.  In the case of the real singlet, this symmetry allows the singlet be a viable dark matter candidate~\cite{McDonald:1993ex,Burgess:2000yq,BahatTreidel:2006kx,Barger:2007im,Davoudiasl:2004be}.
\item[(b)] Requiring that $V(H,\cs)$ possess a global $U(1)$ symmetry eliminates all terms in Eq.~(\ref{eq:pot}) having complex coefficients (e.g. the $\delta_{1},\delta_{3},a_{1},b_{1},c_{1},c_{2},d_{1}$ and $d_{3}$ terms).  
\end{itemize}

In our earlier work on the real scalar SM extension~\cite{Profumo:2007wc,Barger:2007im}, we found that one could generate a strong, first order EWPT and alleviate the direct search-EWPO tension by giving a zero-temperature vev to the real scalar field. As a result, the real scalar mixes with the neutral component of $H$, leading to two unstable mass eigenstates and no dark matter candidate. In the absence of a singlet vev, the real scalar singlet may be a viable CDM candidate but its presence does not affect EWPO.  Moreover, it appears difficult to generate a strong first order EWPT in this case~\cite{Espinosa:2007qk} (see also Refs.~\cite{Anderson:1991zb,Espinosa:1993bs})\footnote{The authors of Ref.~\cite{Espinosa:2007qk} did obtain a strong first order EWPT with the addition of twelve real scalars having no vevs. These scalars contribute to the finite temperature effective potential solely through loop corrections.}. 

In the present case, giving a zero temperature vev to $\cs$ yields a massive scalar $S$ that mixes with the neutral component of $H$ and a massless Goldstone boson $A$ that does not mix. Although the $A$ is, therefore, stable, it is a massless degree of freedom that is not phenomenologically viable, as discussed below\footnote{It also contributes to the number of effective relativistic degrees of freedom and would modify the effective number of light neutrinos in the early universe.}. In order to obtain a viable CDM candidate, we give the $A$ a mass by introducing a soft breaking of the global $U(1)$. We choose the breaking terms that are technically natural and that do not generate additional soft symmetry-breaking terms through renormalization\footnote{For example, the $\delta_{1}$ term in Eq. \ref{eq:pot} can induce 2, 3 and 4-point vertices via SM Higgs loops.}. It is straightforward to see that the $b_1$-term of Eq.~(\ref{eq:pot}) satisfies this requirement. However, retention of only this $U(1)$-breaking term yields a potential having a discrete $\mathbb Z_{2}$ symmetry. To avoid the possibility of cosmological domain walls generated when this symmetry is broken by the vev of $\cs$~\cite{Zeldovich:1974uw,Kibble:1976sj,Kibble:1980mv,Abel:1995wk,Friedland:2002qs}, we include the $U(1)$- and $\mathbb Z_{2}$-breaking linear term proportional to $a_1$ as well.  The resulting potential is
\bea
V_\mathrm{cxSM}&=&\frac{m^{2}}{2}H^{\dagger}H+\frac{\lambda}{4}(H^{\dagger}H)^{2}+\frac{\delta_{2}}{2}H^{\dagger}H|\cs|^{2}+\frac{b_{2}}{2}|\cs|^{2}+\frac{d_{2}}{4}|\cs|^{4}
\label{eq:u1pot} \\
\nonumber
&+&\left(\frac{|b_{1}|}{4}e^{i\phi_{b_1}}  \cs^{2}+|a_1|\, e^{i\phi_{a_1}} \cs  + c.c.\right)
 \ \ \ .
\eea

Depending on the relative sizes of the terms in Eq.~(\ref{eq:u1pot}), we arrive at four distinct phenomenological classes of the complex scalar singlet model.  We summarize these four cases here and in Table \ref{tab:cases}.

\begin{table}[htbp]
\caption{Summary of the four different phenomenological classes allowed by the potential of Eq.~(\ref{eq:u1pot}). Here, $S$ and $A$ denote the real and imaginary components of $\cs$, defined with respect to its vev, $\langle \cs \rangle = v_{S}/\sqrt{2} $. The SM Higgs boson is denoted by $h_{SM}$. The third column denotes the behavior of $V_\mathrm{cxSM}$ under global $U(1)$ symmetry: \lq\lq $U(1)$" indicates $b_1=a_1=0$ while \lq\lq $\slash U(1)$" corresponds to $b_1\not=0$ and (for B2) $a_1\not=0$. The fifth column gives the properties of each scenario relevant to the CDM abundance, while the final column summarizes the potential implications for LHC Higgs studies. } 
\begin{center}
\label{tab:cases}
\begin{tabular}{|c|c|c|c|c|c|}
\hline
Case & Singlet VEV & Symmetry & Masses & Stable states/Pheno & Collider Pheno\\
\hline
A1&$\langle \cs \rangle = 0$ & $U(1)$ & $M_{S}=M_{A}\not=0$ & $S,A$/ identical & $h_{SM}\to SS,AA$\\
A2&$\langle \cs\rangle = 0$ & $\slash{U(1)}$ & $M_{S,A}\not=0$ & $S,A$ & $h_{SM}\to SS,AA$\\
\hline
B1&$\langle \cs\rangle = v_{S}/\sqrt{2} $ & $U(1)$ & $M_{S}\not=0$, $M_{A}=0$ & $A$ & $h_{SM}$-$S$ mixing, $H_{1,2}\to AA$\\
B2&$\langle \cs\rangle = v_{S}/\sqrt{2} $ & $\slash{U(1)}$ & $M_{S,A}\not=0$ & $A$& $h_{SM}$-$S$ mixing, $H_{1,2}\to AA$\\
\hline
\end{tabular}
\end{center}
\end{table}

\newpage
\noindent{\bf{Case A1:}}

The first case imposes a global $U(1)$ symmetry ($a_1=b_1=0$) and does not allow the singlet field to obtain a vev.  In this case, two fields corresponding to the real ($S$) and imaginary ($A$) degrees of freedom of $\cs$ are  degenerate due to the global $U(1)$.  The phenomenology is similar to the real singlet case studied in Refs.~\cite{McDonald:1993ex,Burgess:2000yq,BahatTreidel:2006kx,Barger:2007im,Davoudiasl:2004be}, except that an internal charge is assigned to the singlet field.  The singlet field becomes stable and is then a viable dark matter candidate.  The associated effects of the singlet on the Higgs sector in collider searches for the SM Higgs boson are relevant for this case~\cite{Barger:2007im}.

\noindent{\bf{Case A2:}}

In addition to the $U(1)$ conserving potential, we study the more general non conserving cases. One possibility is that  $\langle \cs\rangle =0$, in which case we require $a_1=0$ while keeping $b_1\not=0$. While $V_\mathrm{cxSM}$ is $\mathbb Z_{2}$ symmetric in this case, we encounter no domain wall problem since the discrete symmetry is not broken.

\noindent{\bf{Cases B1,B2:}}

Here, the singlet obtains a vev.  As a consequence, the field $S$ is allowed to mix with the SM Higgs field.  The resulting effects in the Higgs sector have been studied in detail in the xSM~\cite{Profumo:2007wc,Barger:2007im,BahatTreidel:2006kx,Cerdeno:2006ha,Dedes:2008bf}.  These effects are also generically found in other more complex models that predict a scalar singlet, such as the class of singlet extended supersymmetric models~\cite{Barger:2006dh,Barger:2006rd,Barger:2006sk,Barger:2007ay,Chang:2005ht,Demir:2005kg,Han:1999jc,Miller:2003ay,Panagiotakopoulos:1999ah,Panagiotakopoulos:2000wp,Ellwanger:2003jt} and Randall-Sundrum models where the radion-Higgs mixing is essentially equivalent to singlet-Higgs mixing~\cite{Cheung:2003fz,Csaki:2000zn}.  The field $A$ does not mix with the SM Higgs field in the $U(1)$ symmetric scenario (B1);  it will in general do so for the $U(1)$-breaking scenarios (B2) unless the CP-violating interactions are absent.\footnote{It is possible that the presence of such interactions that are non-vanishing during but not immediately after the EWPT could affect the phase transition dynamics and CDM relic density. We suspect, however, that the impact on the relic density would be minimal since the dark matter (DM) freeze out temperature is typically well below that of the EWPT.} Note that the spontaneously-broken $U(1)$ symmetry of case (B1) yields a massless Goldstone boson that could yield a relic warm or cold dark matter density. The presence of a stable, massive pseudo-Goldstone boson of case (B2) requires $b_1\not=0\not=a_1$ as discussed above. When treating this case in detail below, we can without loss of generality redefine the phase of the complex singlet by $S\rightarrow Se^{i(\pi-\phi_{a_{1}})}$, which is equivalent to taking $\phi_{a_{1}}=\pi$.\footnote{The phenomenology of both cases B1 and B2 can also be obtained more generally in other versions and parameter ranges of the singlet models, such as in a Higgs-portal model~\cite{Patt:2006fw} with the hidden sector symmetry being $\mathbb{Z}_{2}$}

\noindent{\bf Global Minima:}

For the potential to have a global minimum we require that the potential is bounded below and that there are no flat directions.  In what follows, we will take $\lambda > 0$ and $d_2>0$ while allowing $\delta_2$ to range over positive and negative values. When $\delta_2$ is positive, the potential is bounded and there exist no flat directions. For $\delta_2 < 0$, these requirements give the following restrictions on the quartic parameters
\be
\label{eq:boundedness}
\lambda > 0, \quad\quad d_{2} > 0,\quad\quad \lambda d_{2} > \delta_{2}^{2}.
\ee

It is convenient to represent the complex singlet as $\cs = \left[x+iy\right]/\sqrt{2}$ and $H=h/\sqrt{2}$ to obtain the minimization conditions of the potential.  We will always take $\phi_{b_{1}}=\pi$.  This allows us to impose simple conditions which ensure that the vev of $y$ is zero, i.e. there is no mixing between the scalar and pseudoscalar mass eigenstates, as will be shown in Appendix~\ref{appendix:global}.  This in turn implies that CP is not violated (The CP-violating case is not considered in the present study.).  In the special case $a_{1}=0$ the $\phi_{b_{1}}=\pi$ condition is without loss of generality.  With these simplifications we can write the minimization conditions of the potential as
\bea
\label{eq:mincond1}
\frac{\partial V}{\partial h}&=&\frac{h}{2} \left(m^{2}+\frac{\lambda h^{2}}{2}+\frac{\delta_{2}(x^{2}+y^{2})}{2} \right)=0\ \ \ ,\\
\label{eq:mincond2}
\frac{\partial V}{\partial x}&=&\frac{x}{2} \left(b_{2}-|b_{1}|+\frac{\delta_{2} h^{2}}{2}+\frac{d_{2}(x^{2}+y^{2})}{2} \right) - \sqrt{2}|a_{1}|=0\ \ \ ,\\
\label{eq:mincond3}
\frac{\partial V}{\partial y}&=&\frac{y}{2} \left(b_{2}+|b_{1}|+\frac{\delta_{2} h^{2}}{2}+\frac{d_{2}(x^{2}+y^{2})}{2} \right)=0\ \ \ .
\label{eq:micond}
\eea

\noindent These conditions allow four solutions, two of which allow the SM Higgs to accommodate electroweak symmetry breaking: $\langle H\rangle\slash{=}0$ and either $\langle \cs\rangle=0$ or $\langle \cs\rangle\slash{=}0$.  We next obtain the conditions under which these two cases arise:

\noindent{\bf{Vanishing singlet vev:}}

As discussed above, this scenario requires $a_1=0$. In order to guarantee that the extremum at $(v\not=0,v_{S}=0)$ is the global minimum, we must ensure that (a) the eigenvalues of $M^2_\mathrm{scalar}$, are positive and (b) either a secondary minimum with $v_{S}\not=0$ cannot occur or if it does that it is not the global minimum. The first requirement is satisfied when $m^2<0$ and 
\be
\label{eq:detpos}
\delta_2 v^2/2+b_2 > |b_1|\ \ \ .
\ee

This requirement is easily seen from the form of
\be
\label{eq:msquarednovev}
 M^2_\mathrm{scalar} = \mathrm{diag}\, \left(M_h^2,M_S^2, M_A^2\right) \ \ \ ,
 \ee
after eliminating $m^{2}$ in terms of $v$ in the $\{h,x,y\}$ basis, where   
\bea
\label{eq:novev1}
M_{h}^{2}&=&\half \lambda v^{2} \ \ \ ,\\
\label{eq:novev2}
M_{S}^{2}&=&-\half |b_{1}| + \half b_{2}+\frac{\delta_{2}v^{2}}{4} \ \ \ ,\\
\label{eq:novev3}
M_{A}^{2}&=&\half |b_{1}| + \half b_{2}+\frac{\delta_{2}v^{2}}{4} \ \ \ .
\eea

The conditions under which requirement (b) is satisfied are derived in Appendix \ref{appendix:global}. 

\noindent{\bf{Spontaneously broken $U(1)$:}}

For this scenario, we take $a_1\not=0$ to avoid the possibility of domain walls, and we eliminate $m^2$ and $b_2$ in terms of $v$, $v_{S}$ and the other parameters in the potential. With the aforementioned choice of phases, the singlet vev is purely real\footnote{The details of this result are explained in Appendix \ref{appendix:global}}. The resulting mass-squared matrix for the fluctuations about the vevs is
\be
\label{eq:vevmass}
M^2_\mathrm{scalar} = \left(
\begin{array}{ccc}
\lambda v^2/2 & \delta_2 v v_{S}/2 & 0 \\
 \delta_2 v v_{S}/2 & d_2 v_{S}^2/2+\sqrt{2} |a_{1}|/v_{S} & 0 \\
 0 & 0 & |b_{1}| +\sqrt{2} |a_{1}|/v_{S}
\end{array}
\right)\ \ \ ,
\ee

We again require positive eigenvalues of the mass-squared matrix for fluctuations around the point $(v\not=0,v_{S}\not=0)$, leading to
\be
\label{eq:positivity2}
\lambda v^{2}+d_{2}v_{S}^{2}+\frac{2\sqrt{2}|a_{1}|}{v_{S}}>\sqrt{(\lambda v^{2}-d_{2}v_{S}^{2}-\frac{2\sqrt{2}|a_{1}|}{v_{S}})^{2}+4\delta_{2}^{2}v^{2}v_{S}^{2}} \ \ \ .
\ee

This condition is simplified to
\be
\label{eq:positivity3}
\lambda(d_{2}+\frac{2\sqrt{2}|a_{1}|}{v_{S}^{3}})>\delta_{2}^{2} \ \ \ .
\ee
Using the methods described in Appendix \ref{appendix:global} we find that there are no other conditions are needed to ensure that this point in the global minimum.

The aforementioned conditions can be relaxed if the minimum is a metastable local minimum rather than a global minimum. Although we do not consider this possibility here, we note that a viable, metastable minimum must be one with a sufficiently long lifetime and one into which the universe initially cools. We refer the reader to Refs.~\cite{Barger:2003rs,Kusenko:1996xt} and references therein for further details.

\section{Scalar Sector Spectra and Couplings}
\label{sec:spectra}

The different symmetry-breaking scenarios outlined above lead to distinct spectra for the scalar sector of the cxSM. Here, we delineate the various possibilities.

\subsection{Vanishing Singlet VEV}

In the case of a vanishing singlet vev, for which we set $a_1=0$, the minimization conditions in Eq.~(\ref{eq:mincond1}) can be used to relate the scalar masses to the parameters $\lambda$, $b_1$, $b_2$,  and $v$.  The mass-squared matrix $M^2_\mathrm{scalar}$ is given by Eqs.~(\ref{eq:msquarednovev}-\ref{eq:novev3}). In this case none of the neutral scalars mix with each other, and we obtain a two-component dark matter scenario. Moreover,  if we set the $U(1)$ breaking parameter, $b_{1}$,  to zero, $M_S=M_A$ due to the restored $U(1)$ symmetry.  In the limit of a small $b_{1}$, the singlet mass splitting parameter 
\be
\label{eq:masssplit}
\Delta \equiv \left|{M_{A}-M_{S}\over M_{A}+M_{S}}\right|\approx \left|{ b_{1}\over 2b_{2}+ \delta_{2}v^{2}}\right| \ \ \ ,
\ee

provides a useful handle on the contribution of $S$ to the total CDM relic density.  For large values of $d_2$ the annihilation process $AA\to SS$ will reduce the density of $A$ after $S$ freezes out in the early universe unless $\Delta$ is small so $A$ and $S$ freeze out at nearly the same time.  Hence only for $\Delta\ll 0.1$ does $\Omega_\mathrm{DM}$ receive significant contributions from annihilations of $S$ (c.f. Fig.~\ref{fig:contrib}).

\subsection{Singlet VEV}

When the singlet field obtains a vev, the masses are given by the eigenvalues of $M^2_\mathrm{scalar}$ in Eq.~(\ref{eq:vevmass}). For the $U(1)$ symmetric potential, $M^2_\mathrm{scalar}$ has one vanishing eigenvalue, corresponding to the Goldstone boson of the spontaneously broken global symmetry. The remaining real scalars $H_{1,2}$ are mixtures of $H$ and $S$, and neither is stable. Because we are interested in the possibility of scalar dark matter, we will not consider this case in detail and concentrate instead on the situation in which the global $U(1)$ is both spontaneously and explicitly broken. From Eq.~(\ref{eq:masquared}), we note that the parameters $b_1$ and $a_1$ give a mass to the $A$. As discussed above, because we have taken these parameters to be real and $M_A^2>0$, the $A$ remains stable and is a candidate for scalar dark matter. The corresponding masses and mixing angles are given by
\bea
M_{H_{1}}^{2}&=&\frac{\lambda v^{2}}{4}+\frac{d_{2}v_{S}^{2}}{4}+\frac{\sqrt{2}|a_{1}|}{2v_{S}}-\sqrt{\left(\frac{\lambda v^{2}}{4}-\frac{d_{2}v_{S}^{2}}{4}-\frac{\sqrt{2}|a_{1}|}{2v_{S}}\right)^{2}+\frac{\delta_{2}^{2}v^{2}v_{S}^{2}}{4}} \ \ \ ,\\
M_{H_{2}}^{2}&=&\frac{\lambda v^{2}}{4}+\frac{d_{2}v_{S}^{2}}{4}+\frac{\sqrt{2}|a_{1}|}{2v_{S}}+\sqrt{\left(\frac{\lambda v^{2}}{4}-\frac{d_{2}v_{S}^{2}}{4}-\frac{\sqrt{2}|a_{1}|}{2v_{S}}\right)^{2}+\frac{\delta_{2}^{2}v^{2}v_{S}^{2}}{4}} \ \ \ ,\\
\label{eq:masquared}
M_{A}^{2}&=&|b_{1}|+\frac{\sqrt{2}|a_{1}|}{v_{S}} \ \ \ ,\\
\label{eqn:mixingangle}
\tan 2 \phi&=&{ \delta_2 z \over \half \lambda - \half d_2 z^2 - 
{\sqrt 2 |a_1| \over v^3 z} } \ \ \ ,
\eea

where $z=v_{S}/v$ is the relative size of the singlet vev.

\subsection{Annihilation and the Relic Density}

\begin{figure}[htbp]
\begin{center}
\includegraphics[angle=0,width=0.12\textwidth]{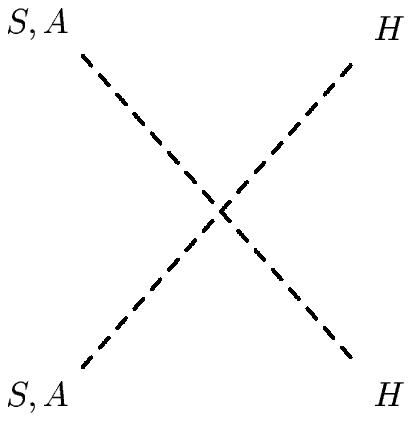}~~~
\includegraphics[angle=0,width=0.12\textwidth]{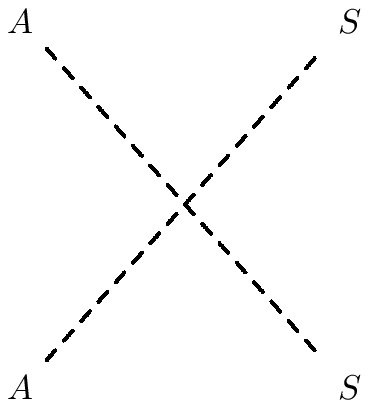}~~~
\includegraphics[angle=0,width=0.15\textwidth]{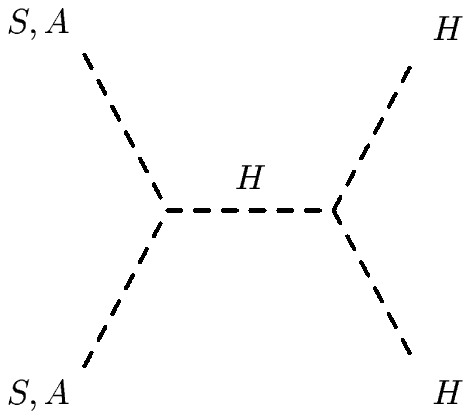}~~~
\includegraphics[angle=0,width=0.12\textwidth]{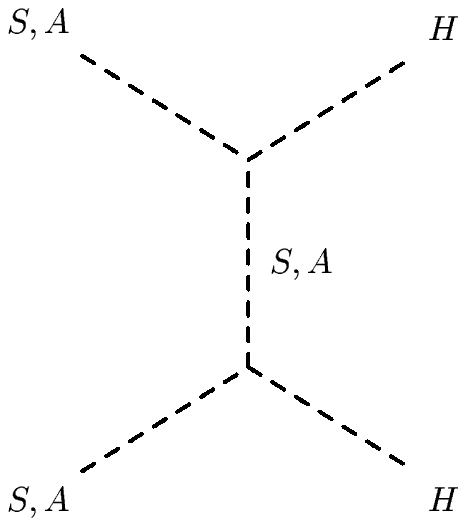}~~~
\includegraphics[angle=0,width=0.15\textwidth]{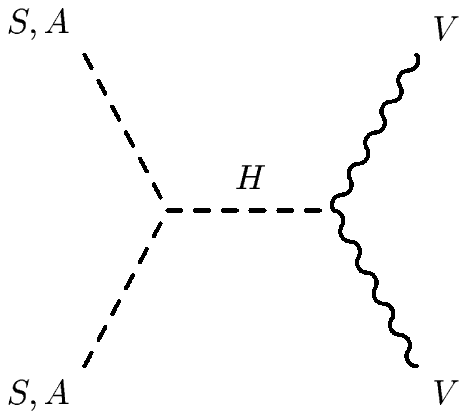}~~~
\includegraphics[angle=0,width=0.15\textwidth]{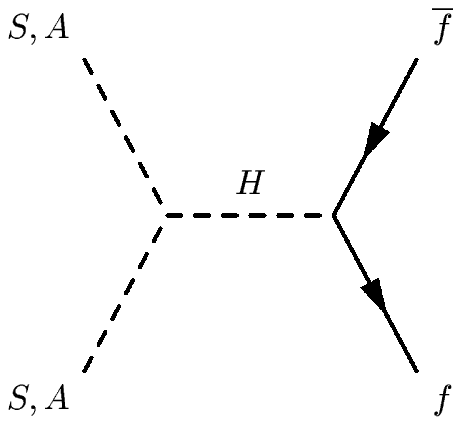}
\caption{Annihilation processes that contribute to the thermally averaged cross section for the two-component scalar DM scenario ($v_{S}=0$).  Here, $H$ is the SM Higgs boson, $f$ is a SM fermion, and $V$ is any of the SM gauge bosons. The fields $S$ and $A$ are quanta created by the real and imaginary parts of $\cs$, respectively.}
\label{fig:annFDcaseA}
\end{center}
\end{figure}

When $v_{S}=0$, annihilation processes involving both the $S$ and the $A$ are important for determining the CDM relic density. 
The Feynman diagrams for this case, shown in Fig. \ref{fig:annFDcaseA}, are similar to those for dark matter composed of a single, real scalar singlet.  The difference in the present instance is that two singlets appear, and when the magnitude of $\Delta$ is relatively small, contributions from both species can have a significant impact on the relic density. 

\begin{figure}[htbp]
\begin{center}
\includegraphics[angle=0,width=0.16\textwidth]{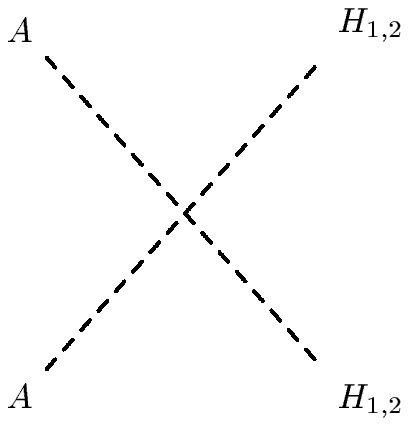}~~~
\includegraphics[angle=0,width=0.16\textwidth]{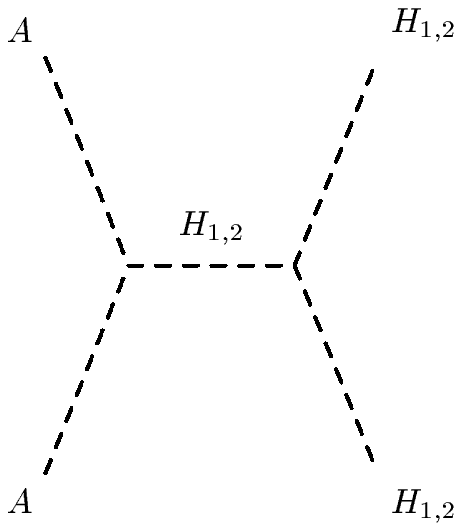}~~~
\includegraphics[angle=0,width=0.16\textwidth]{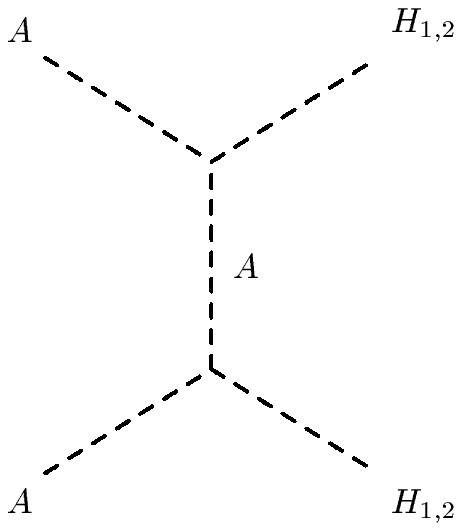}~~~
\includegraphics[angle=0,width=0.19\textwidth]{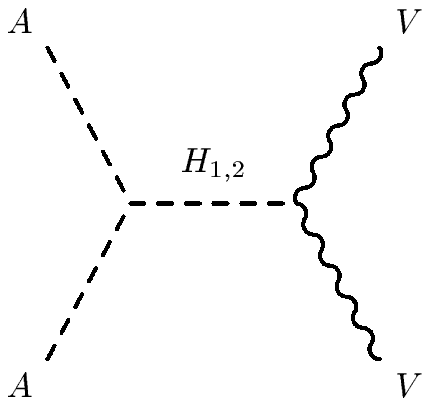}~~~
\includegraphics[angle=0,width=0.19\textwidth]{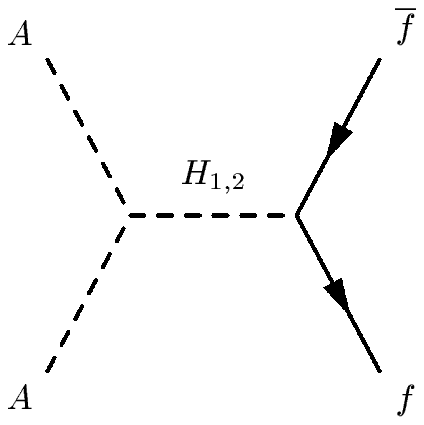}
\caption{Annihilation processes that contribute to the thermally averaged cross section for the case of the singlet vev.  All processes are mediated via the two Higgs eigenstates. The notation is as in Fig. \ref{fig:annFDcaseA}, except that $H_j$ ($j=1,2$) denote the two unstable neutral scalars.}
\label{fig:annFDcaseB}
\end{center}
\end{figure}

When $v_{S}\not=0$, only the $A$ is stable. One must take now into account the presence of two massive, unstable scalars $H_{1,2}$ into which pairs of $A$ scalars may annihilate: $AA\leftrightarrow H_i H_j$ with $i$ and $j$ running over the labels 1 and 2. The corresponding diagrams are shown in Fig. \ref{fig:annFDcaseB}.

\section{Constraints}
\label{sec:const}

When analyzing the collider and dark matter phenomenology of the cxSM, we consider a number of constraints implied by direct searches for new scalars, electroweak precision data, and astrophysical observations.

\subsection{Collider constraints}

\ben
\item The LEP-II experiments constrain the $ZZ\phi$ coupling for light $\phi$~\cite{Sopczak:2005mc}.  If the mass of the scalar field is below 114 GeV, the coupling must be reduced below the SM Higgs coupling to $Z$ bosons, which may be achieved in this model through singlet-Higgs mixing.  Such mixing is only present within the complex singlet model if the singlet obtains a nonzero VEV.  

\item The limit on new physics contributions to the invisible $Z$ width is 1.9 MeV at 95\% C.L.~\cite{Amsler:2008zz,Barger:2006dh}.  The contribution, from the decays $Z\to Z^{*} H^{*}\to \nu \bar \nu S S$ and $\nu \bar \nu AA$ are many orders of magnitude below this limit.  

\item The mixing of the neutral $SU(2)$ and singlet scalars affect electroweak precision observables  (EWPO) through changes in the gauge boson propagators.  Since EWPO favor a light SM Higgs boson, any singlet that is sufficiently mixed with the SM Higgs boson is also favored to be relatively light~\cite{Barger:2007im,Profumo:2007wc}.

\item For a very light state that mixes with the SM Higgs field, the amount of mixing can be severely limited by experimental limits on $B \to H_i X$ and $\Upsilon\to H_i \gamma$ decays~\cite{O'Connell:2006wi,Dermisek:2006py}.  The mass ranges for the lightest Higgs state we consider do not go into the region where these constraints are relevant.
\een

\subsection{Astrophysical constraints}

\ben
\item One of the most rigorous constraints that can be applied to these models are the limits from the WMAP 5-year survey and spatial distribution of galaxies on the relic density of DM~\cite{Komatsu:2008hk}
\be
\Omega_{\rm DM} h^2=0.1143\pm 0.0034 \ \ \ ,
\ee
provided the cxSM contribution to the relic density does not exceed this bound.  The Hubble constant is $h=0.701\pm 0.013$.  The subsequent results for the relic density we provide are calculated using Micromegas 2.0~\cite{Belanger:2006is}.

\item The limits on the elastic cross section from DM scattering off nuclear targets have considerably improved in the last few years.  Present limits from XENON 10kg~\cite{Angle:2007uj} and the CDMS five-tower~\cite{Ahmed:2008eu} experiment are the most stringent spin-independent scattering with a lowest upper bound of $~4\times 10^{-8}$ pb.  The Super Kamiokande experiment~\cite{Habig:2001ei,Desai:2004pq} places a bound on the spin-independent and spin-dependent scattering cross-sections of order $10^{-5}$ pb and $10^{-2}$ pb, respectively.  Scalar DM predicts a vanishing spin-dependent elastic cross section.  

\item If the present baryon asymmetry in the universe has an electroweak origin, the singlet may aide in ensuring a sufficiently strong first order EWPT.  We do not rigorously apply constraints from this sector on the parameters, but observe  that the presence of the  quartic interaction $H^\dag H |\cs|^2$  can lead to the requisite phase transition provided that the coupling $\delta_2$ 
is negative~\cite{Profumo:2007wc}. We discuss this region of parameter space and the corresponding implications for the DM relic density below. 

\item Observations of the Bullet cluster may be used to place a constraint on the quartic DM coupling\footnote{For massive DM the quartic coupling is the singlet parameter $d_{2}$.  However, for massless DM the quartic coupling appears through radiative loops.}.  Accordingly, the DM scattering cross section over the DM mass must be less than $1.25$ cm$^{2}$/g~\cite{Randall:2007ph}.  Using similar methods as Refs.~(\cite{Bento:2000ah,McDonald:2007ka}) we obtain the following constraint on the DM mass and quartic coupling $g_{DM}$
\be
\frac{M_{DM}^{3}}{g_{DM}^{2}} > 2\times 10^{-5} \ \text{GeV}^{3} \ \ \ .
\ee
If we take the quartic couplings to be $g_{DM}=\sqrt{4\pi}$ this constraint only excludes $M_{DM} < 64$ MeV, which is well below the range of DM mass that we study.  In particular, the scenario with a massless, self-interacting scalar [case (B1) of Table \ref{tab:cases}] is ruled out.
\een

\section{Scalar Sector Phenomenology}
\label{sec:pheno}

We now detail the phenomenological consequences of the four classes of the complex singlet sector.

\subsection{Two Component DM: $U(1)$ symmetric scenario}

If the $U(1)$ symmetry is imposed, the cxSM is equivalent to a model with two real singlets of the same mass and internal charge assignment.  Much of the phenomenology is similar to that of the xSM as discussed in Refs.~\cite{McDonald:1993ex,Burgess:2000yq,O'Connell:2006wi,BahatTreidel:2006kx,Barger:2007im,He:2007tt,Davoudiasl:2004be} and elsewhere. For example,  the two real singlets couple to the SM via their interactions with the Higgs boson, and they can play important roles in Higgs searches.  Specifically, the branching fractions of Higgs boson decays to SM fields may be reduced due to dominant decays to singlet pairs, resulting in large missing energy in the events.  If the decay to a singlet pair (or \lq\lq invisible decay") is allowed, the usual SM search modes would have a substantially reduced likelihood for observing a signal.  However, the Higgs decay to invisible states channel may itself be a promising search mode for the SM Higgs boson~\cite{Eboli:2000ze,Davoudiasl:2004aj,atlas:2003ab}.

In addition to its impact on collider searches, the stable singlet can serve as a viable DM candidate that correctly reproduces the relic density yet evades present direct detection bounds~\cite{Barger:2007im}.  

\subsection{Two Component DM: Explicit $U(1)$ breaking}

\begin{figure}[htbp]
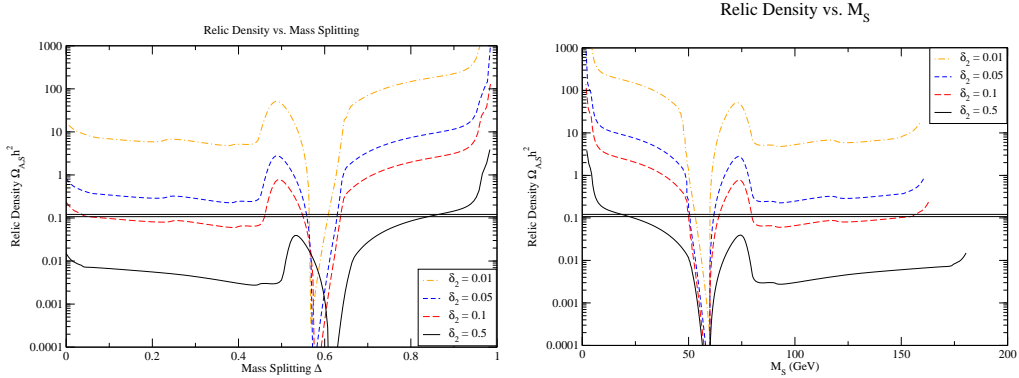

\begin{center}
\includegraphics[width=0.40\textwidth]{figs/rel-spl.eps}\hspace{0.01\textwidth}
\includegraphics[width=0.40\textwidth]{figs/rel-ma.eps}
\caption{Relic density variation with the mass splitting parameter $\Delta$.  We show a few illustrations with the singlet-higgs coupling parameter $\delta_{2}=0.01,0.05,0.1$ and $0.5$.  With the choices of parameters we made each curve corresponds to a constant sum of the singlet masses squared: $M_{S}^{2}+M_{A}^{2}=b_{2}+\delta_{2}v^{2}/2$.}
\label{fig:1brd}
\end{center}
\end{figure}

Explicit breaking of the $U(1)$ symmetry forces the singlet masses to split, with the size of splitting dependent on the magnitude of the symmetry breaking parameter, $b_{1}$ (recall that $a_1$ must vanish when $v_{S}=0$).  Both states are stable, with the lightest being the DM candidate. In the early universe, the heavier state annihilates efficiently into the lighter state for large values of $d_{2}$ effectively eliminating it unless $\Delta$ is small. For small values of $d_{2}$ the contribution from the heavier state to the overall relic density will depend more on annihilations to SM particles.  In the limit that $b_{1}\to 0$, both states annihilate equally and freeze out at the same time resulting in a relic density that is double the case with only one real singlet. This effect is shown in Fig. \ref{fig:1brd} for $M_{H}=120$ GeV, $b_{2}=50000\text{ GeV}^{2}$, and $d_{2}=1$.  In the relic density plot we can easily see the Higgs pole which occurs at a DM mass of $60$ GeV.  Other features can be seen in the relic density as we get dips when new annihilation channels open\footnote{New annihilation channels open when $M_{S}$ increases.  For our scan in Fig. \ref{fig:1brd} and Eq. \ref{eq:masssplit} this corresponds to decreasing mass splitting.}, increasing the annihilation cross section and thus decreasing the relic density.  In particular, one must have $\delta_{2}\gtrsim 0.1$ in order to avoid overproducing the relic CDM density, except for DM masses in the vicinity of the Higgs pole.

\begin{figure}[htbp]
\begin{center}
\includegraphics[width=0.6\textwidth]{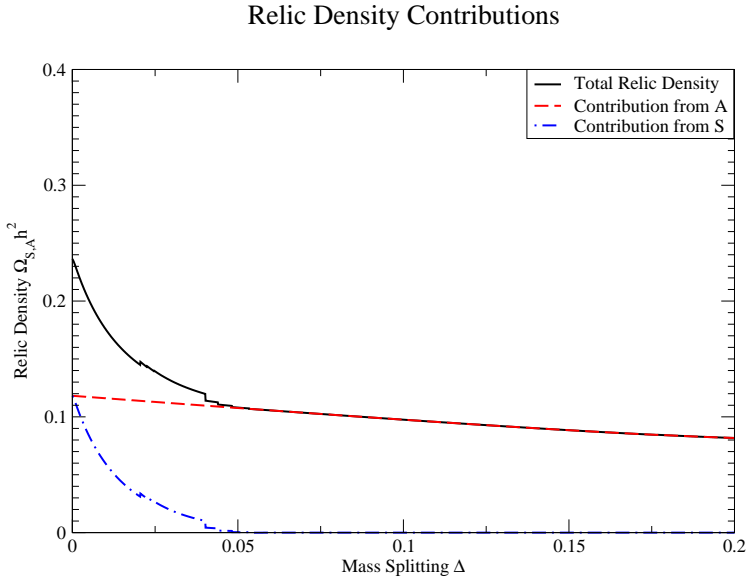}
\caption{Relic density contributions from both singlet particles.  At low mass splitting there are contributions from both $S$ and $A$ to the total relic abundance.}
\label{fig:contrib}
\end{center}
\end{figure}

In Fig. \ref{fig:contrib} we show the contributions of the two singlets to the total relic abundance.  As the mass splitting approaches zero we get a significant  contribution from the heavier singlet $A$.  In the limit that the mass splitting approaches zero the relic density from $A$ is the same as that from $S$. The doubling of the relic density at $\Delta=0$ can be attributed to summing over the two $U(1)$ charges. 

\begin{figure}[htbp]
\begin{center}
\includegraphics[width=0.6\textwidth]{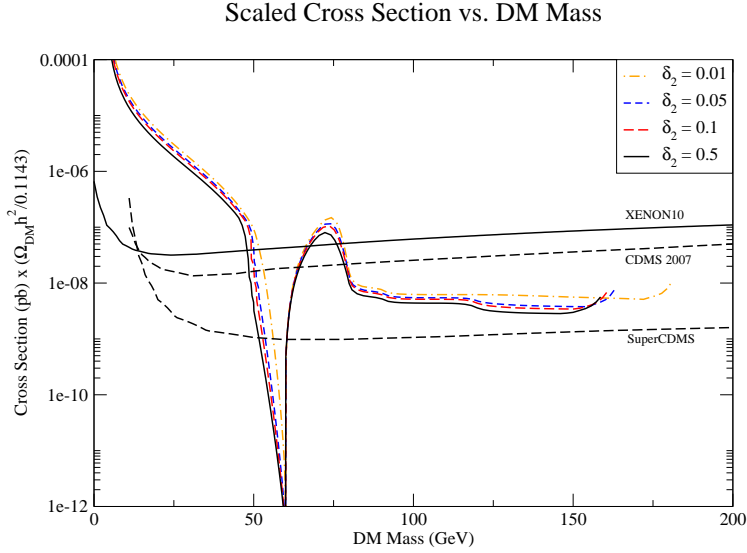}
\caption{Elastic scattering cross section off proton targets for the curves shown in Fig. \ref{fig:1brd}, appropriately scaled to the relic density.  Direct detection curves from current and future experiments are also displayed.}
\label{fig:1bdd}
\end{center}
\end{figure}

In Fig. \ref{fig:1bdd}, we show the predicted direct detection rates from DM-proton elastic scattering for the values of $\delta_{2}=0.01,0.05,0.1$ and $0.5$.  The $\delta_{2}$ parameter influences the $M_{S}$, $M_{A}$ mass splitting and the couplings among the Higgs and singlets.  The cross sections in Fig. \ref{fig:1bdd} are scaled with the calculated relic density relative to that measured by the WMAP 5-year result~\cite{Komatsu:2008hk} in order to properly compare the predicted cross sections with those given by direct detection experiments, which present their results assuming the observed density.  Because of this scaling the scattering cross section closely follows the relic density.  Note that current direct detection limits exclude DM masses below $\sim M_{H}/2$ for all values of $\delta_{2}$ assuming that the scattering cross section scales with the relic density $\sigma_{S-p}\to \sigma_{S-p}\times(\Omega_{DM}h^{2}/0.1143)$.

The impact of the two stable states in this model on Higgs searches at the LHC is pronounced, but not radically different than the case of one real singlet.  If the singlets are light enough to allow the decays $h\to AA$ and/or $h\to SS$, the branching fractions of the Higgs boson to SM particles is reduced to
\be
\text{BF}(H \to X_{SM}) =\text{BF}(h_{SM} \to X_{SM}){ \Gamma_{h_{SM}}\over  \Gamma_{h_{SM}}+\Gamma({H \to SS})+\Gamma({H \to AA})} \ \ \ ,
\ee

where the partial widths to singlet pairs are given by
\be
\Gamma(H\to SS) = {g_{HSS}^{2} \over 32 \pi M_H}\sqrt{1-{4M_S^2\over M_H^2}}, \quad\quad\Gamma(H\to AA) = {g_{HAA}^{2} \over 32 \pi M_H}\sqrt{1-{4M_A^2\over M_H^2}} \ \ \ .
\ee

The $U(1)$ breaking does not change the interaction between the singlet fields and the Higgs so the $HSS$ and $HAA$ couplings are identical
\be
g_{HSS}=g_{HAA}=-\half \delta_{2}v \ \ \ .
\ee

The only difference between the relative decay rates is due to the different masses.

\subsection{Massless DM: Spontaneous $U(1)$ breaking}

As discussed above, spontaneous breaking of the global $U(1)$ symmetry leads to a massless Goldstone boson.  Such a massless propagating mode has severe constraints from big bang nucleosynthesis\footnote{Though it is possible to avoid big bang nucleosynthesis constraints, as shown in Ref.~\cite{Cerdeno:2006ha}}~\cite{Barger:2003zg,Cyburt:2004yc} and the Bullet cluster, as discussed in Section \ref{sec:const}.

\subsection{Single Component DM \& the EWPT: Spontaneous and soft $U(1)$ breaking}

The scenario allowing for the richest array of physics possibilities for both cosmology and collider phenomenology involves the simultaneous spontaneous and explicit breaking of the global $U(1)$ symmetry.  As indicated earlier, one obtains two massive, unstable scalars ($H_{1,2}$) that involve mixtures of the $h_{SM}$ and $S$ and one massive stable scalar ($A$) that can contribute to the CDM relic density. Moreover, the presence of a non-vanishing singlet vev at zero temperature has been shown to allow for a strong, first order EWPT as needed for successful electroweak baryogenesis under appropriate conditions for the doublet-singlet interaction terms in the potential, $V_\mathrm{cxSM}$~\cite{Profumo:2007wc}. 

Beginning with the collider implications, we recall that the production cross-sections of these scalar states may be smaller than the corresponding SM Higgs.  The signal reduction factor of a traditional Higgs decay mode $X_{SM}$ is
\be
\xi_{i}^{2} = R_{i1}^{2}\times \text{BF}(H_{i}\to X_{\text{SM}})\ \ \ ,
\label{eq:xi}
\ee

where 
\be
R_{i1}=\left\{\begin{array}{c} \cos\phi \quad\quad i=1 \\\sin\phi \quad\quad i=2 \end{array}\right. \ \ \ ,
\ee

is the SM Higgs component of the state $i$ and the mixing angle $\phi$ is given in Eq. \ref{eqn:mixingangle}.  In practice, we may take the magnitude of $a_1$ such that $|a_1| \ll d_2 (zv)^3 = d_2 v_{S}^3$ while still avoiding the presence of domain walls.  For either small $\delta_{2}$ or small $z$, the mixing angle $\phi$ can be small, and the predictions are hard to distinguish from the SM.  The proper decay length of the dominantly singlet scalar can be comparable to the size of the detector for very small $\phi$, but the mixing would need to be less than $10^{-6}$ to produce observable displaced vertices~\cite{Barger:2007im}.  For even smaller mixing, the singlet may be a metastable state which can complicate the relic density constraints as it could account for a fraction of the dark matter in the universe today.  In this extreme case the mixing to place the lifetime of the singlet state comparable to the age of the universe is $|\phi| \lesssim 10^{-21}$.

The combination of mixing between the SM Higgs and the CP-even singlet along with decays to invisible CP-odd singlet states can make Higgs searches at the LHC challenging.  This scenario mimics the nearly minimal supersymmetric standard model~\cite{Dedes:2000jp,Panagiotakopoulos:1999ah,Panagiotakopoulos:2000wp} of the singlet extended supersymmetric model~\cite{Menon:2004wv,Barger:2006dh,Barger:2006kt,Barger:2006sk,Balazs:2007pf}, with a tadpole singlet term in the superpotential.  In the nearly minimal supersymmetric standard model the singlet-Higgs mixing and the Higgs decays to invisible singlino dominated neutralino states reduce the discovery potential of the Higgs boson at the LHC.  

\begin{figure}[htbp]
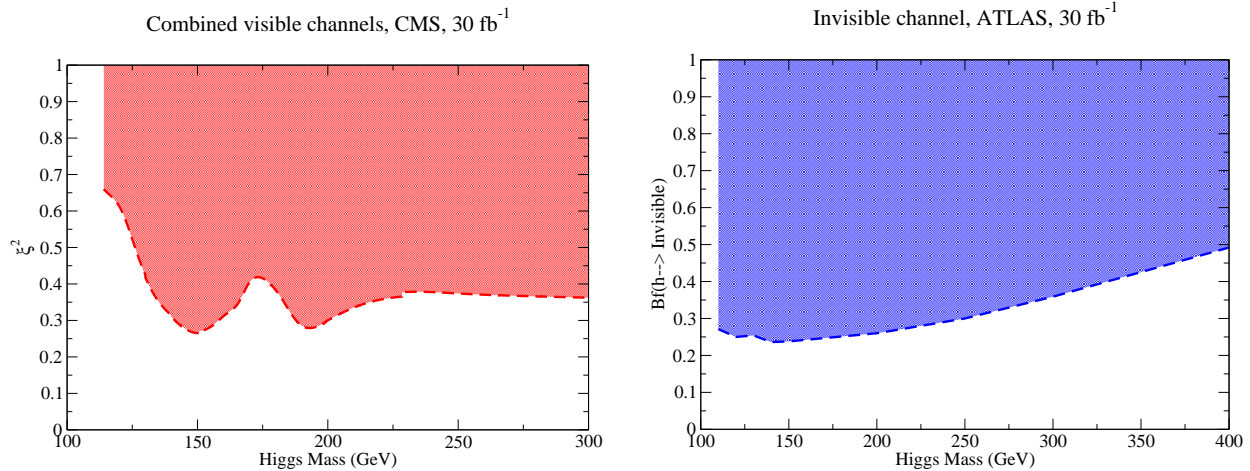

\begin{center}
\includegraphics[width=0.48\textwidth]{figs/xi-redfact.eps}\hspace{0.03\textwidth}
\includegraphics[width=0.48\textwidth]{figs/invbf-reach.eps}
\caption{Mixing parameter ranges (shaded) that can yield $5\sigma$ discovery of a scalar boson through (a) traditional Higgs discovery modes at CMS for 30 fb$^{-1}$ of data and (b) invisible SM Higgs search modes at ATLAS with 30 fb$^{-1}$ of data.  In part (a) the quantity $\xi^{2}$ is defined in Eq. \ref{eq:xi}.}
\label{fig:hdisc}
\end{center}
\end{figure}

Using the expected significance of the CMS and ATLAS detectors for detection of a SM Higgs signal, we can estimate the range of Higgs-singlet mixing that would allow at least a $5\sigma$ significance discovery with 30 fb$^{-1}$ of data\footnote{This is based on scaling the significance of the Higgs signal with the reduction factor, $\xi^{2}$, at the LHC for 30 fb$^{-1}$ given in the CMS TDR~\cite{cmstdr}. We have not included effects of systematic uncertainties.}.  In Fig. \ref{fig:hdisc}a, we show that region from visible channels for the CMS experiment, given in terms of $\xi^{2}$, the signal reduction factor defined in Eq. (\ref{eq:xi}).  

The LHC also has sensitivity to a Higgs boson that decays to states which escape without detection~\cite{Eboli:2000ze}, which in our case is the massive stable scalar $A$.  If the Higgs boson is produced in weak boson fusion, the sign of missing $p_T$ and the azimuthal correlation of the forward jets can allow the signal to be extracted from the QCD and electroweak $W,Z+jj$ background.  The reach for a SM Higgs boson decaying to invisible states is given at the ATLAS detector with 30 fb$^{-1}$ of integrated luminosity as the shaded region in Fig. \ref{fig:hdisc}b~\cite{atlas:2003ab}.  If mixing is present, the invisible branching fraction reach is weakened.  The minimum invisible branching fraction (BF) needed for a 5$\sigma$ discovery must be larger than for a pure SM-like Higgs scalar as
\be
\text{BF}_{\mathrm{min}}(H_i \to \text{Invisible}) = \frac{1}{R_{i1}^{2}}\text{BF}_{\mathrm{min}}(h_{SM}\to \text{Invisible}) \ \ \ ,
\label{eq:bfreduce}
\ee

where the $R_{i1}^{2}$ term reflects the change in production strength of $H_{i}$.

\begin{figure}[htbp]
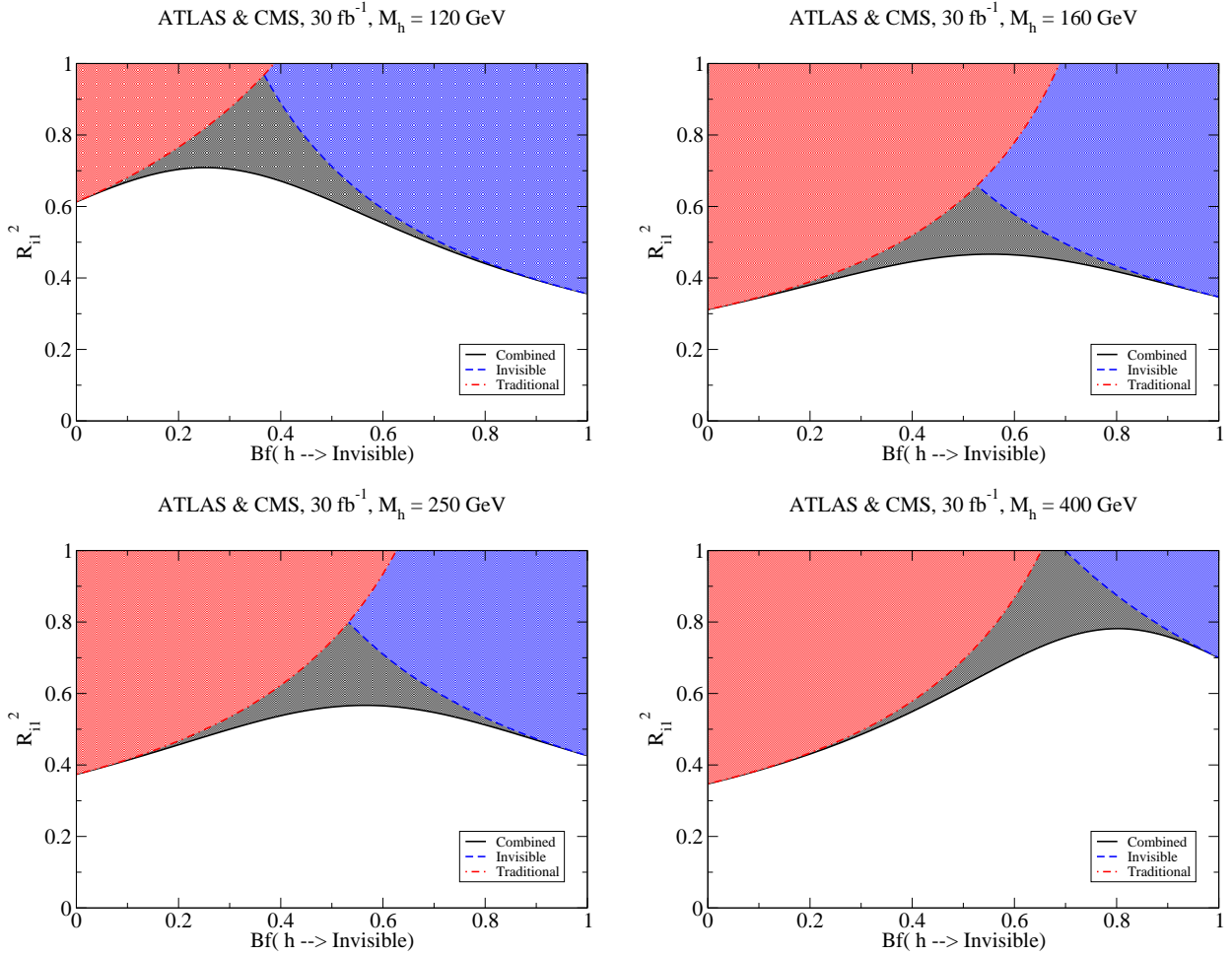

\begin{center}
\includegraphics[width=0.48\textwidth]{figs/reach-mh120.eps}\hspace{0.03\textwidth}
\includegraphics[width=0.48\textwidth]{figs/reach-mh160.eps}\\
\vspace{.1in}
\includegraphics[width=0.48\textwidth]{figs/reach-mh250.eps}\hspace{0.03\textwidth}
\includegraphics[width=0.48\textwidth]{figs/reach-mh400.eps}
\caption{Higgs boson $5\sigma$ discovery potential at the LHC at 30 fb$^{-1}$ through the traditional Higgs search for $m_{h}=120, 160, 250$ and 400 GeV at CMS (coverage indicated by the red shaded region) and the search at ATLAS via the invisible decay modes (blue region).  We also show the improvement from combining visible and invisible limits (gray region).}
\label{fig:mix-invdec}
\end{center}
\end{figure}

To estimate the reach at the LHC when both visible and invisible decays can occur, we show in Fig. \ref{fig:mix-invdec} the Higgs boson discovery potential at the LHC at $30$ fb$^{-1}$ for the visible Higgs search at CMS and for the search at ATLAS via the invisible decay modes with Higgs state masses $M_{h}=120, 160, 250$ and $400$ GeV.  We also show the combined search limit.  In combining the search limits, we do not take into account systematic effects that may be dominant at small signals. 

The two Higgs mass eigenstates have complementary SM Higgs fractions, with the smallest possible value of the Higgs fraction of the SM-like state being $R_{i1}^2 = \half$.  Thus from Fig. \ref{fig:mix-invdec} the prospects for observing at least one Higgs boson are good for $M_{h}\lesssim 160$ GeV since $R_{i1}^{2}=\half$ is contained entirely within the three discovery regions.  

When additional non-SM Higgs decays are possible, the statistical significance of the Higgs signal in traditional modes is given by Eq. \ref{eq:xi}, which can be re-expressed as
\bea
\xi_i^2 &=& R_{i1}^2 \left[1-\text{BF}(H_i \to \Delta)\right] \ \ \ ,
\label{eq:xiopenmode}
\eea

\noindent where $\Delta$ represents any state that is not a SM decay mode of the Higgs boson.

\begin{figure}[htbp]
\begin{center}
\includegraphics[width=0.47\textwidth]{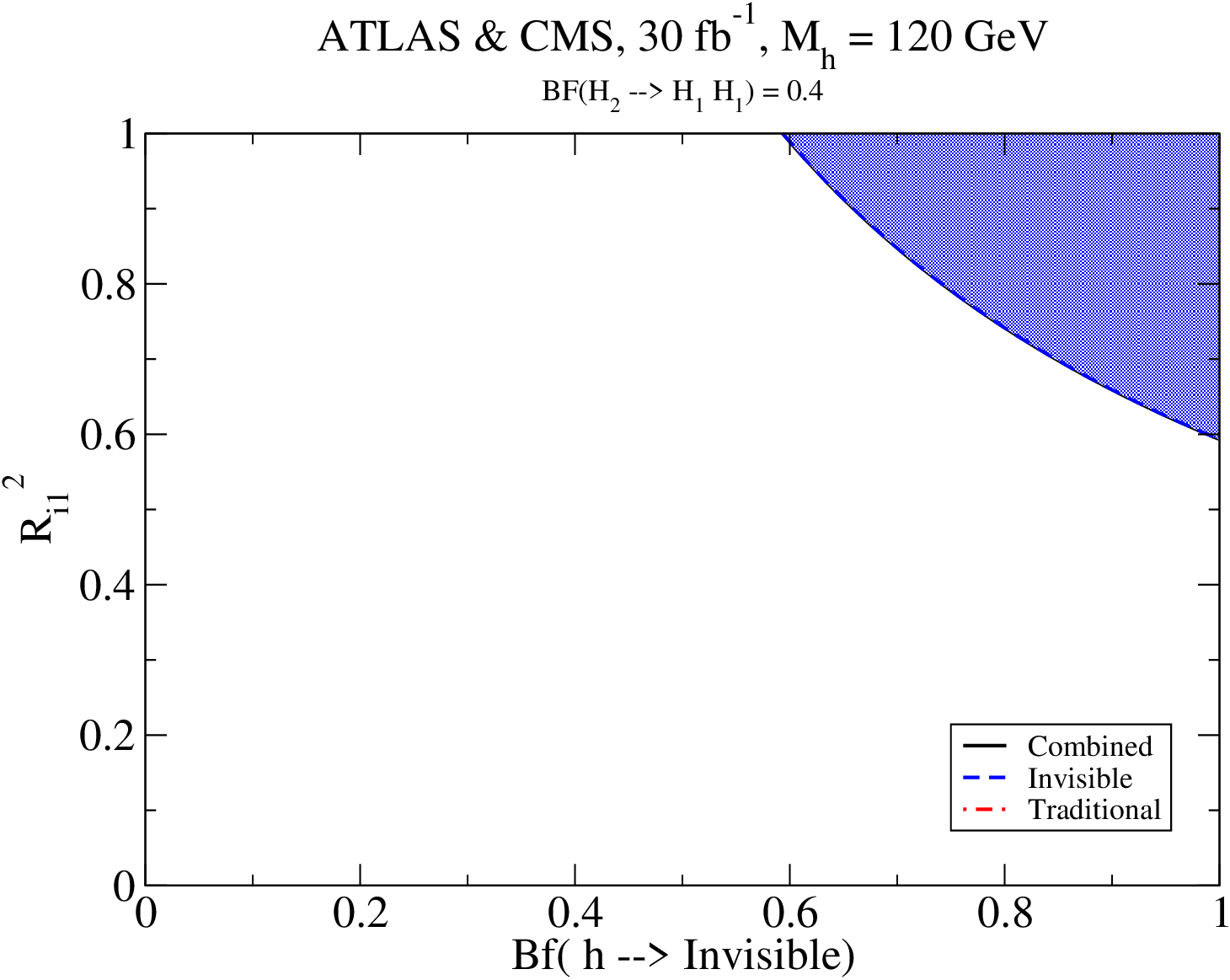}
\hspace{0.03\textwidth}
\includegraphics[width=0.47\textwidth]{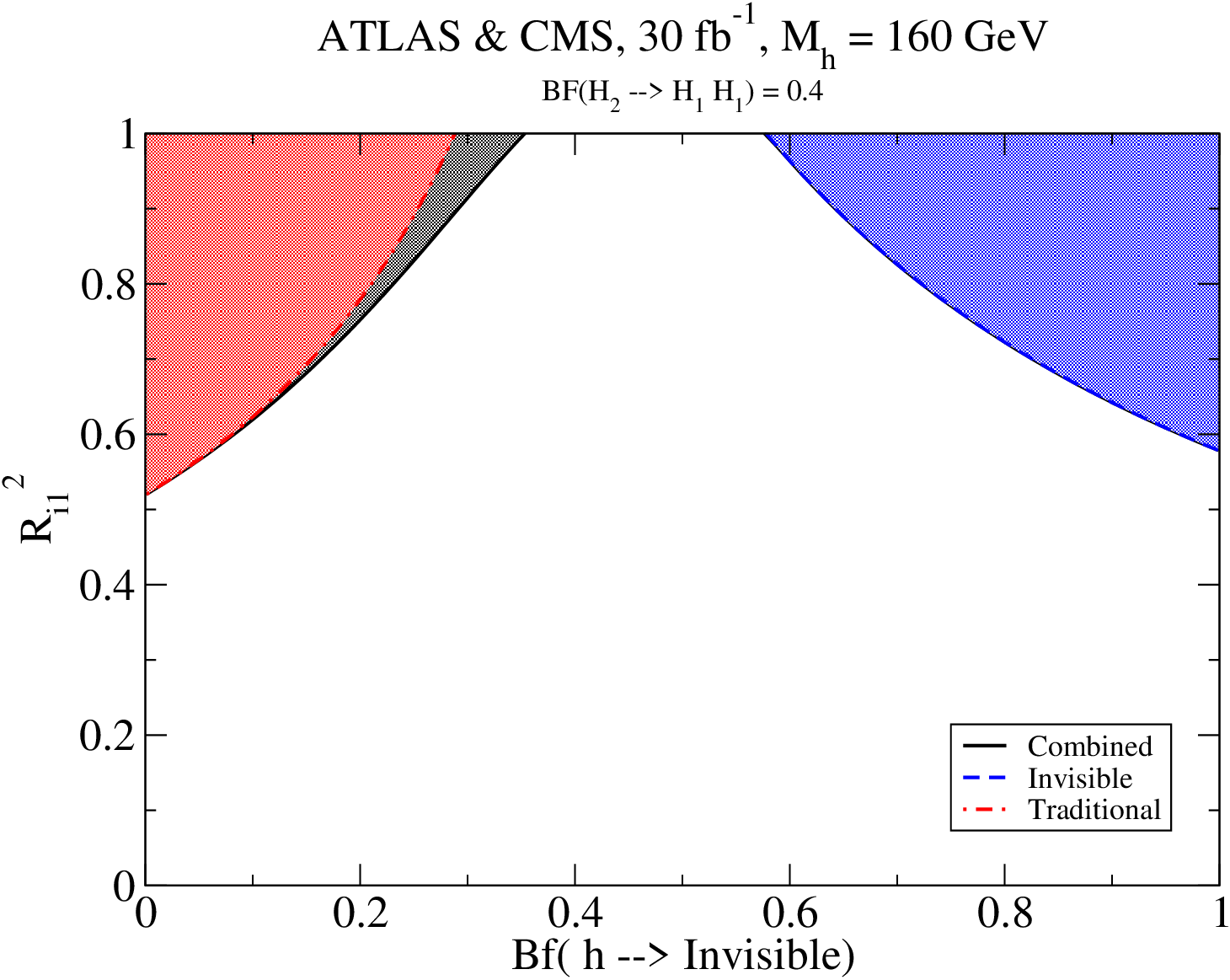}\\
\vspace{.1in}
\includegraphics[width=0.47\textwidth]{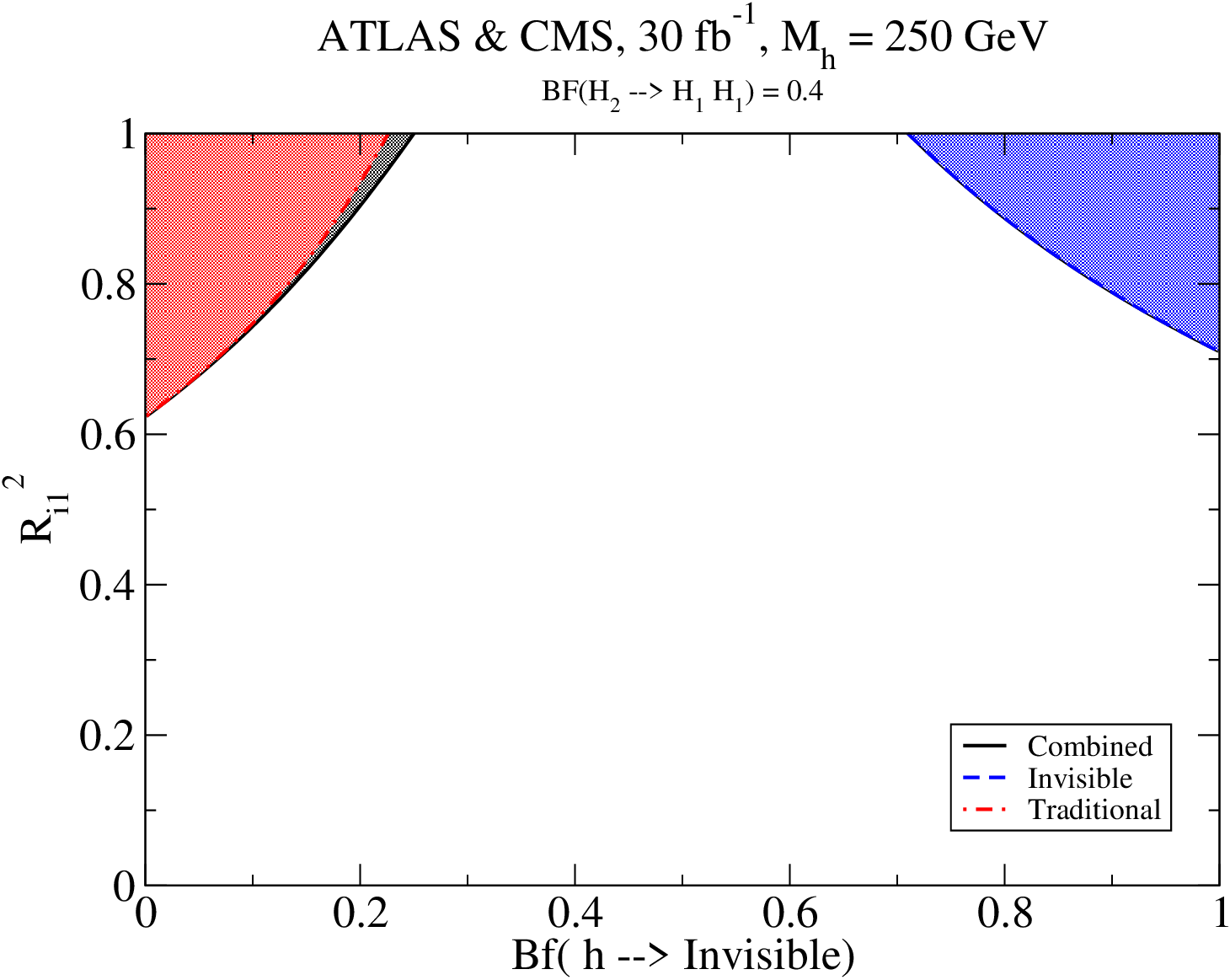}
\hspace{0.03\textwidth}
\includegraphics[width=0.47\textwidth]{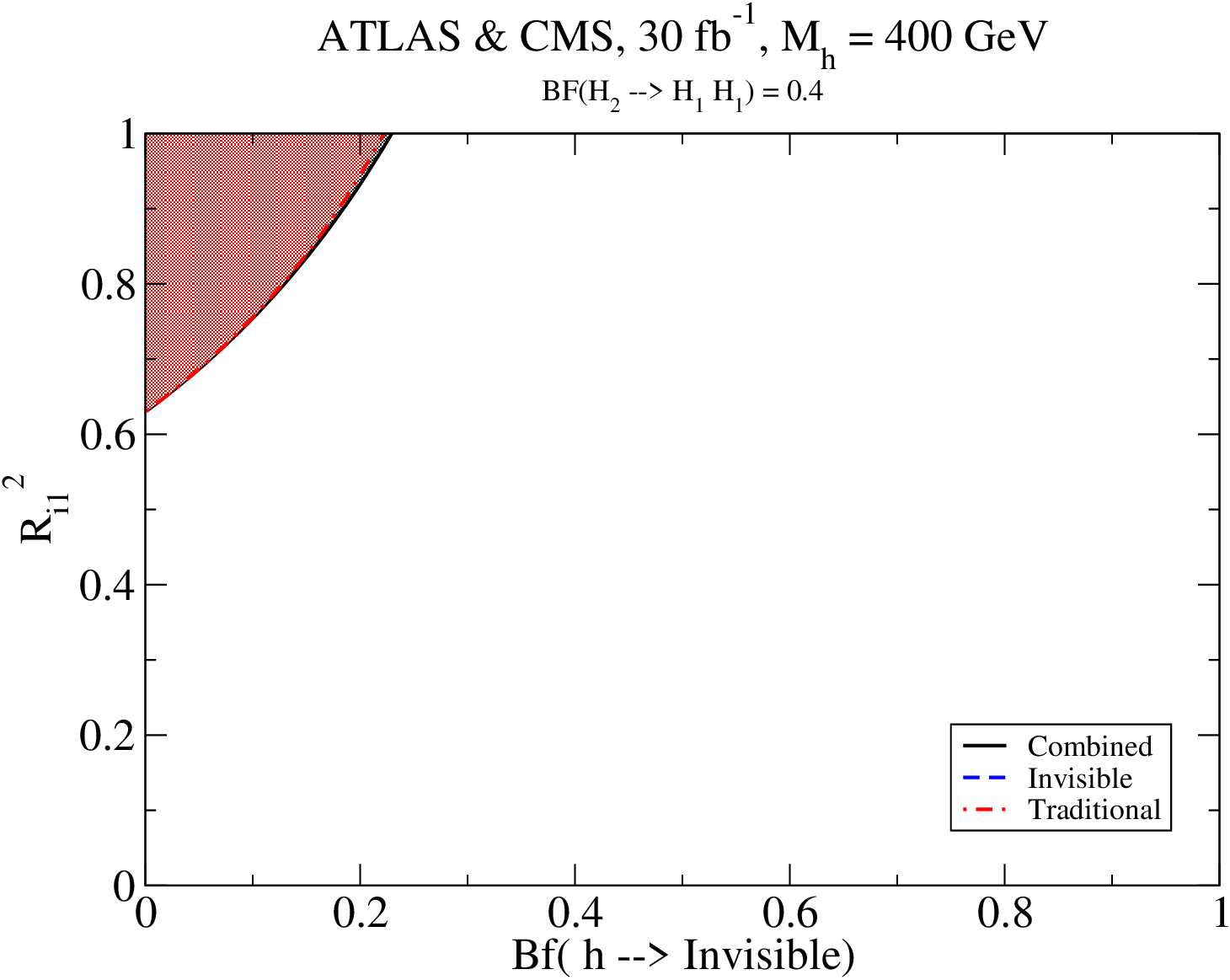}
\caption{Same as Fig. \ref{fig:mix-invdec} except with $\text{BF}
(H_2\to H_1 H_1) = 0.4$.}
\label{fig:invdecwsplit}
\end{center}
\end{figure}

For example, when the $H_2\to H_1 H_1 \to 4 X_{SM}$ decay mode is detectable and visible, the reach from the traditional search modes will be reduced by the signal reduction factor in Eq. \ref{eq:xiopenmode}.  For similar reasons, the invisible decay reach of Eq. \ref{eq:bfreduce} will be altered to
\be
{\text{BF}_{\mathrm min}(h_{SM}\to \text{Invisible})\over R^2_{i1} (1-\text{BF}(H_2\to H_1 H_1))} \ \ \ ,
\ee

\noindent where the new term describes the decrease of the effective strength of producing an invisible decay due to the additional decay $H_2\to H_1H_1$\footnote{Note that this reduction can be more generally applied to any other decay mode which steals the signal from $H\to X_{SM}$ or $H\to Invisible$ such as $H\to 6j$ in $R$-parity violating SUSY models~\cite{Carpenter:2008sy}, or displaced Higgs decays in Hidden Valley models~\cite{Strassler:2006ri}.}.  To illustrate this effect, we show in Fig. \ref{fig:invdecwsplit} the reach of the traditional search and invisible search combined if the splitting decay occurs with a branching fraction of 40\%.

The presence of the Higgs splitting mode does reduce the Higgs discovery potential in this example drastically.  However, it is expected that with higher luminosity over the $30$ fb$^{-1}$ assumed here, discovery via visible modes may still possible.  Further, discovery via the Higgs splitting mode itself is also an interesting alternative in such cases~\cite{Carena:2007jk,Chang:2008cw}.

\begin{figure}[htbp]
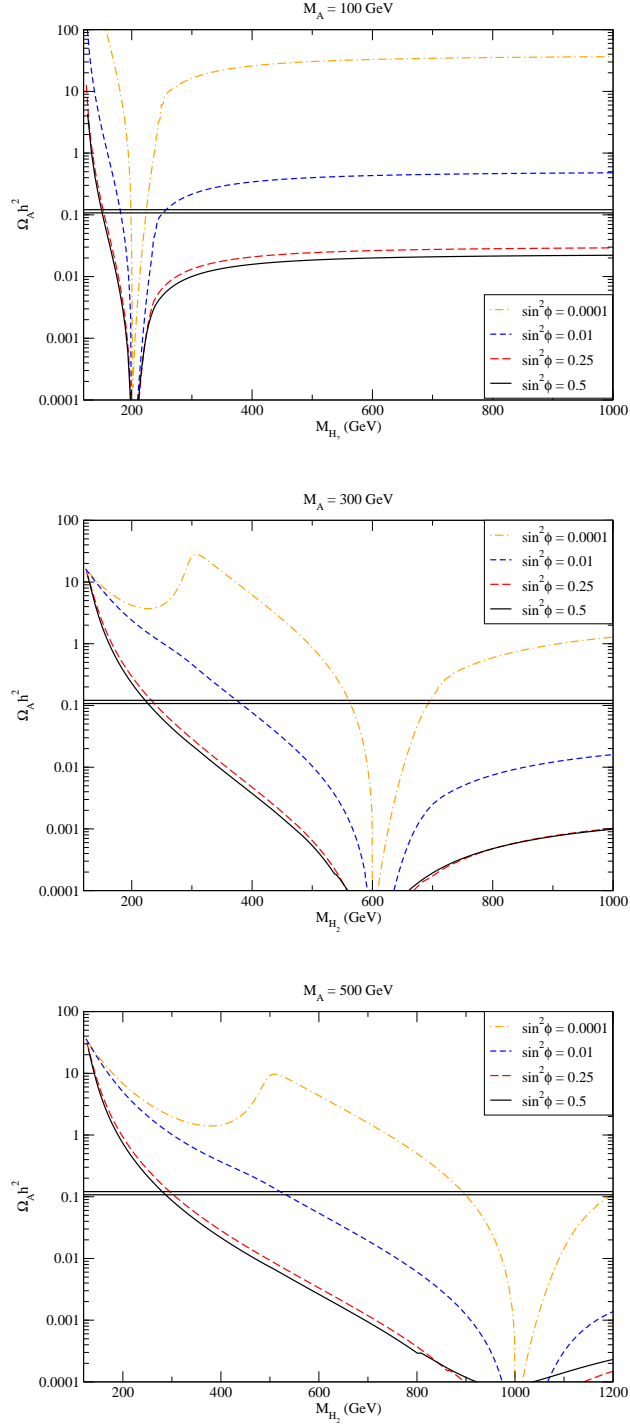

\begin{center}
\includegraphics[width=0.5\textwidth]{figs/MA100.eps}\\
\vspace{.25in}
\includegraphics[width=0.5\textwidth]{figs/MA300.eps}\\
\vspace{.25in}
\includegraphics[width=0.5\textwidth]{figs/MA500.eps}
\caption{Scan over the heavier Higgs eigenstate mass with the lighter Higgs mass fixed at $120$ GeV. The DM mass and mixing between the Higgs eigenstates are fixed at selected values.  For all three values of $M_{A}$ the effect of the $H_{2}$ pole is evident.}
\label{fig:rel-d2scan}
\end{center}
\end{figure}

To study the dark matter phenomenology of this model we set the parameters $\lambda$, $\delta_{2}$, and $b_{1}$ so that the lightest Higgs eigenstate has a mass of 120 GeV, the mixing between the Higgs eigenstates is fixed at a few selected values, and the dark matter mass is fixed.  We then scan over the only remaining free parameter $d_{2}$ which effectively is a scan over the mass of the heavier Higgs eigenstate.  The results of these scans are shown in Fig. \ref{fig:rel-d2scan}.  The results point to an interesting connection with EWPO, which favor the presence of an additional light neutral scalar that mixes strongly with the neutral $SU(2)$ scalar.  The analysis of Refs.~\cite{Barger:2007im,Profumo:2007wc} suggests that the region with $M_{H_{2}}\lesssim 200$ GeV is favored when $\sin 2\phi$ is nearly maximal.  Combining these considerations with the results from Fig. \ref{fig:rel-d2scan}, we observe that the scenarios with relatively light scalar DM would be favored, as they allow for $M_{H_{2}}\lesssim 200$ GeV and large mixing without overproducing the relic density.

An interesting feature of this scenario is the relation between the parameter $\delta_2$ that governs the relic density and the strength of the EWPT. In order to prevent the washout of the baryon asymmetry produced during the phase transition, one must satisfy the inequality
\be
\label{eq:ewpt1}
\frac{v(T_C)}{T_C} \gtrsim 1 \ \ \ ,
\ee

where $T_C$ is the critical temperature and $v(T)$ is the $SU(2)_L$ vev at temperature $T$. It has long been known that the LEP II direct search bounds on the SM Higgs mass preclude the SM scalar sector from satisfying this inequality. However, it was shown in Ref.~\cite{Profumo:2007wc} that the addition of one real, singlet scalar to the SM Higgs sector could satisfy Eq.(\ref{eq:ewpt1}) while yielding a SM Higgs with mass greater than 114.4 GeV. Since the cxSM analyzed here shares many features with the real singlet scalar extension of Ref.~\cite{Profumo:2007wc} (the \lq\lq xSM"), we refer to the results of that study in order to evaluate the prospects for a strong first order EWPT. In doing so, we note the similarities and differences between the cxSM and the xSM:

\begin{itemize}
\item[(i)] The real singlet scenario also includes cubic terms in the zero temperature potential (before spontaneous symmetry-breaking) that we have not included here. As we discuss below, however, these cubic terms are not essential ingredients for a scalar extension that satisfies inequality (\ref{eq:ewpt1}). 
\item[(ii)] The quartic interaction $\delta_2 (H^\dag H) |\cs|^2/2$ includes a quartic coupling $(H^\dag H) S^2$ that also appears in the xSM and that was shown in Ref.~\cite{Profumo:2007wc} to drive a strong first order EWPT for appropriate values of the coupling, even in the absence of cubic interactions between the singlet and SM doublet. In what follows, we amplify on this point. 
\item[(iii)] The presence of the additional degree of freedom with the complex scalar (the $A$) that becomes the dark matter scalar will also generate an additional one-loop contribution to the finite temperature effective potential  beyond those included in the analysis of Ref.~\cite{Profumo:2007wc}. We do not anticipate this addition to have a significant impact on the finite temperature analysis, as the dominant effect of the new scalars are primarily via the tree-level terms in the potential. 
\end{itemize}

Having these features in mind, we now discuss the prospects for a first order EWPT in the cxSM universe. 
Assuming that $\langle \cs \rangle =0$ for $T > T_C$, Eq.~(\ref{eq:ewpt1}) implies that~\cite{Profumo:2007wc}
\be
\label{eq:ewpt2}
\frac{16 E_\mathrm{SM} }{2{\bar\lambda}_0(T_C) + 4 \delta_2 \tan^2\alpha_C + d_2\tan^4\alpha_C}\gtrsim 1 \ \ \ ,
\ee

where $E_\mathrm{SM}$ is the coefficient of the cubic term in the finite temperature effective potential generated by gauge boson loops, ${\bar\lambda}_0$ is the corresponding coefficient of the quartic power of the $SU(2)_L$ classical field\footnote{The corresponding condition is given in Eq. (4.11) of Ref. \cite{Profumo:2007wc}. Here, we have chosen a different normalization for the Higgs quartic coupling, leading to the factor of 16 rather than four in the numerator.}, and
\be
\label{eq:vevratio}
\tan\alpha_C = \frac{v_{S}(T_C)}{v(T_C)}\ \ \ .
\ee

In the SM, one has $\tan\alpha_C=0$ as there is no singlet contribution to the effective potential. Given the value of $E_\mathrm{SM}$ computed to one-loop order in the SM and the relation between ${\bar\lambda_0}$, $v(T=0)$, and the SM Higgs boson mass $m_H$, one finds that $m_H\lesssim 45$ GeV in order to satisfy the criterion (\ref{eq:ewpt1}). The results of non-perturbative analysis of the effective potential increase this upper bound to roughly 70 GeV~\cite{Aoki:1999fi}. In order to obtain a strong first order EWPT and a mass of the Higgs boson consistent with the LEP II direct search, one may exploit the terms in the denominator of Eq.~(\ref{eq:ewpt2}). In particular, choosing a negative value for $\delta_2$ can reduce magnitude of the denominator, relaxing the requirements on the value of ${\bar\lambda}_0$ that governs the Higgs boson mass. The numerical analysis of  Ref.~\cite{Profumo:2007wc} applied to the SM extension with a single, real singlet scalar indicates that one could satisfy the criterion of Eqs.~(\ref{eq:ewpt1},~\ref{eq:ewpt2}) for $0 > \delta_2 > -1$ and values for $v_{S}$ ranging from a few GeV up to order 100 GeV while satisfying the requirements of the boundedness of the potential and the LEP II lower bound on the Higgs boson mass. 

\begin{figure}[htbp]
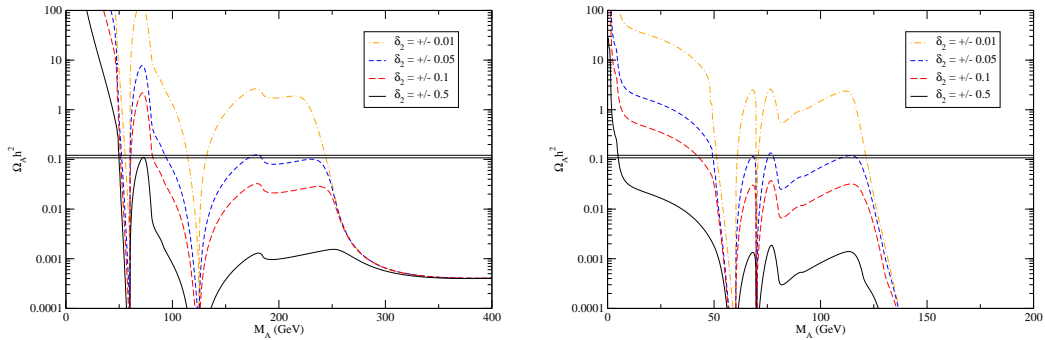

\begin{center}
\includegraphics[width=0.4\textwidth]{figs/vs100.eps}\hspace{0.03\textwidth}
\includegraphics[width=0.4\textwidth]{figs/smallvs.eps}\hspace{0.03\textwidth}
\caption{Relic density as a function of $M_A$ for selected values of the quartic coupling $\delta_2$. The left panel corresponds to the choice $v_{S}=100$ GeV, while $v_{S}=10$ GeV for the right panel.}
\label{fig:rel-vs}
\end{center}
\end{figure}

To analyze the compatibility of this scenario for EWPT with cxSM scalar dark matter, we have computed the relic density for negative values of $\delta_2$. The results are shown in Fig.~\ref{fig:rel-vs}, where we plot $\Omega_\mathrm{DM} h^2$ vs. $M_A$ for selected values of $\delta_2$. The left panel gives the results for $v_{S}=100$ GeV, with $M_{H_{1}}=120$ GeV and $M_{H_{2}}=250$ GeV. The right panel corresponds to a much smaller singlet vev, $v_{S}=10$ GeV, with $M_{H_{1}}=120$ GeV and $M_{H_{2}}=140$ GeV.

In both cases, the dips in the relic density correspond to resonantly-enhanced annihilation rates at $M_{H_{i}}=2 M_A$. Moreover, we find that the sign of $\delta_{2}$ has little effect on the relic density.  The only noticeable difference in the relic density between positive and negative values of $\delta_{2}$ occurred for DM masses between the Higgs masses where interference terms in the process $AA\to H_{i}H_{j}$ would be present, though the effect of these interference terms was very small.  More importantly, the results show that one may obtain the observed relic density in the cxSM for a broad range of values for $v_{S}$ and either sign for $\delta_2$.  These ranges of parameters are consistent with those identified in the Ref.~\cite{Profumo:2007wc} that lead to a strong first order EWPT and a Higgs scalar having a mass above the LEP II direct search bound.  Thus, it appears quite possible that the cxSM will accommodate the observed relic density, the EWPT needed for successful electroweak baryogenesis, and the present constraints from collider searches and electroweak precision data.  

\vspace{0.2in}

\section{Conclusions}
\label{sec:concl}

The cxSM analyzed here presents a rich array of possibilities for addressing outstanding problems at the particle and nuclear physics-cosmology interface.  When the model contains a global $U(1)$ symmetry that is softly but not spontaneously broken, it yields a two-component DM scenario that is relatively simple compared to others discussed recently in the literature~\cite{McDonald:1993ex,Burgess:2000yq,BahatTreidel:2006kx,Barger:2007im,Davoudiasl:2004be}.  When the $U(1)$ symmetry is both spontaneously as well as softly broken, we obtain a single-component DM scenario that also contains the necessary ingredients for a strong, first order EWPT as required for successful electroweak baryogenesis.  When the scalar dark matter is relatively light, the latter scenario also allows for mixing between the real component of the singlet field and the neutral $SU(2)$ scalar without overproduction of the relic density.  This mixing of the real fields can alleviate the tension between direct search lower bounds on the Higgs boson mass and EWPO that favor a light SM-like scalar.

These features of the cxSM would remain academic in the near future if it could not be discovered at the LHC.  Indeed, mixing between the neutral $SU(2)$ and real singlet scalars tends to weaken the LHC discovery potential for a light scalar when only traditional Higgs search modes are considered.  As we have shown above, however, a combination of both these traditional modes and the invisible search channels can allow one to probe nearly all the phenomenologically interesting parameter space of the model in the early phases of the LHC in the absence of significant Higgs splitting decays. Additional luminosity should ultimately allow one to search for the cxSM even when such splitting modes are present.  In short, this scenario appears to be simple, interesting from the standpoint of cosmology, and testable.  As such, its discovery could yield new insights into the puzzles of symmetry breaking in the early universe.

\section{Acknowledgments}

VB and PL thank the Aspen Center for Physics for hospitality.  This work was supported in part by the U.S. Department of Energy under grants No. DE-FG02-95ER40896, DE-FG02-05ER41361, DE-FG02-08ER41531, and Contract DE-AC02-06CH11357, by the Wisconsin Alumni Research Foundation, the Institute for Advanced Study (PL), and by the National Science Foundation grant No. PHY-0503584.

\appendix

\section{Global Minima and Complex Singlet VEVs}
\label{appendix:global}

To find the global minimum of the potential we first need to satisfy the minimization conditions
\bea
\label{eq:min1}
\frac{\partial V}{\partial h}&=&\frac{h}{2} \left(m^{2}+\frac{\lambda h^{2}}{2}+\frac{\delta_{2}(x^{2}+y^{2})}{2} \right)=0\ \ \ ,\\
\label{eq:min2}
\frac{\partial V}{\partial x}&=&\frac{x}{2} \left(b_{2}-|b_{1}|+\frac{\delta_{2} h^{2}}{2}+\frac{d_{2}(x^{2}+y^{2})}{2} \right) - \sqrt{2}|a_{1}|=0\ \ \ ,\\
\label{eq:min3}
\frac{\partial V}{\partial y}&=&\frac{y}{2} \left(b_{2}+|b_{1}|+\frac{\delta_{2} h^{2}}{2}+\frac{d_{2}(x^{2}+y^{2})}{2} \right)=0\ \ \ .
\label{eq:minconditions}
\eea

For each of the cases we are studying, $a_{1}=0$ and $a_{1}\ne0$, there are several solutions apart from the solution that gives the global minimum: $(v_{0}\ne0,v_{x}=0,y=0)$ for $a_{1}=0$ and $(v_{0}\ne0,v_{x}\ne0,y=0)$ for $a_{1}\ne0$.  We first use the minimization conditions to solve for $m^{2}$ for both cases and $b_{2}$ for the $a_{1}\ne0$ case and obtain the following:
\bea
m^{2}_{a_{1}=0}&=&-\frac{\lambda v_{0}^{2}}{2}\\
m^{2}_{a_{1}\ne0}&=&-\frac{\lambda v_{0}^{2}}{2}-\frac{\delta_{2}v_{x}^{2}}{2},\qquad b_{2}=|b_{1}|-\frac{\delta_{2}v_{0}^{2}}{2}-\frac{d_{2}v_{x}^{2}}{2}+\frac{2\sqrt{2}|a_{1}|}{v_{x}}
\eea

We assume for each case that the point under consideration is a minimum so the eigenvalues of the respective mass-matrices are required to be positive (Eq.~\ref{eq:detpos} for $a_{1}=0$ and Eq.~\ref{eq:positivity3} for $a_{1}\ne0$). We use the positivity requirement on the mass-squared matrix eigenvalues to rule out alternative solutions as minima of the potential.

When it is possible for a solution to be a minimum, we can exclude it from being the global minimum by comparing the value of the potential with that conjectured to be the global minimum.  

Following the above method of analysis, there is only one other solution in the $a_{1}=0$ case that could possibly be a minimum, namely, the point $v=0,v_{x}\ne0,y=0$.  If one of the two following conditions fails then this point is not a minimum:
\bea
\delta_{2}v_{0}^{2}&>&4M_{S}^{2} \ \ \ ,\\
\delta_{2}(\delta_{2}v_{0}^{2}-4M_{S}^{2})&>&\lambda d_{2}v_{0}^{2} \ \ \ .
\eea
If both of these conditions are satisfied then the point $(v=0,v_{x}\ne0,y=0)$ is a minimum.  It is not the global minimum if
\be
\lambda d_{2}v_{0}^{4}>(\delta_{2}v_{0}^{2}-4M_{S}^{2})^{2} \ \ \ .
\ee

In the $a_{1}\ne0$ case, all possible alternative solutions either have a negative eigenvalue of the mass-squared matrix or is not the global minimum. Hence, for $a_{1}\ne0$ no extra conditions are needed to ensure that the point $(v_{0}\ne0,v_{x}\ne0,y=0)$ is the global minimum.

\bibliographystyle{h-physrev}
\bibliography{csxsm.bib}
\newpage

\end{document}